\documentclass[prX,twocolumn]{revtex4}
\usepackage{amssymb,multirow,color,natbib,amsmath}
\usepackage[dvips]{graphicx}

\renewcommand{\vec}[1]{\mathbf{#1}}

\renewcommand{\(}{\left(}
\renewcommand{\)}{\right)}

\begin{document}
\title{On the energy spectrum of strong magnetohydrodynamic
  turbulence} \author{Jean Carlos Perez$^{1,2}$, Joanne Mason$^3$,
  Stanislav Boldyrev$^2$, Fausto Cattaneo$^3$}
\affiliation{${~}^1$Space Science Center, University of New Hampshire,
  Durham, NH 03824 \\ ${~}^2$Department of Physics, University of
  Wisconsin at Madison, 1150 University Ave, Madison, WI 53706
  \\ ${~}^3$Department of Astronomy and Astrophysics, University of
  Chicago, 5640 S. Ellis Ave, Chicago, IL 60637\\ {\sf
    jcperez@wisc.edu}, {\sf jmason@flash.uchicago.edu}, {\sf
    boldyrev@wisc.edu}, {\sf cattaneo@flash.uchicago.edu} }
\date{\today}

\begin{abstract}
The energy spectrum of magnetohydrodynamic turbulence attracts
interest due to its fundamental importance and its relevance for
interpreting astrophysical data.  Here we present measurements of the
energy spectra from a series of high-resolution direct numerical
simulations of MHD turbulence with a strong guide field and for
increasing Reynolds number. The presented simulations, with numerical
resolutions up to $2048^3$ mesh points and statistics accumulated over
30 to 150 eddy turnover times, constitute, to the best of our
knowledge, the largest statistical sample of steady state MHD
turbulence to date. We study both the balanced case, where the
energies associated with Alfv\'en modes propagating in opposite
directions along the guide field, $E^+(k_\perp)$ and $E^-(k_\perp)$,
are equal, and the imbalanced case where the energies are different.
In the balanced case, we find that the energy spectrum converges to a
power law with exponent $-3/2$ as the Reynolds number is increased,
consistent with phenomenological models that include scale-dependent
dynamic alignment. 
For the imbalanced case, with $E^+>E^-$, the simulations
show that $E^-\propto k_{\perp}^{-3/2}$ for all Reynolds numbers
considered, while $E^+$ has a slightly steeper spectrum at small
$Re$. As the Reynolds number increases, $E^+$ flattens. Since $E^\pm$
are pinned at the dissipation scale and anchored at the driving
scales, we postulate that at sufficiently high $Re$ the spectra will
become parallel in the inertial range and scale as $E^+ \propto
E^-\propto k_{\perp}^{-3/2}$. Questions regarding the universality of 
the spectrum and the value of the ``Kolmogorov constant'' are discussed. 
\end{abstract}


\maketitle

\section{Introduction}
Astrophysical plasmas are typically magnetized and turbulent, with
turbulent fluctuations spanning a tremendous range of scales in which
the energy spectrum follows a power law
scaling~\citep[e.g.,][]{armstrong_rs95,Goldstein_rm95}. Incompressible
magnetohydrodynamics (MHD) provides the simplest theoretical framework
for studying magnetized plasma turbulence. The precise form of the MHD
turbulence spectrum is crucial for a variety of processes in
astrophysical systems with extended inertial intervals, such as plasma
heating and wave-particle interactions, which are sensitive to small
variations in the spatial scaling of the fluctuations
\citep[e.g.,][]{chandran_etal09,chandran_etal10,ng_b10}.  The
incompressible MHD equations take the form
\begin{eqnarray}
 \label{mhd-elsasser}
  \left( \frac{\partial}{\partial t}\mp\vec V_A\cdot\nabla \right)
  \vec z^\pm+\left(\vec z^\mp\cdot\nabla\right)\vec z^\pm &=& -\nabla
  P +\nu\nabla^2 \vec z^{\pm}+\vec f^\pm, \nonumber \\ \nabla \cdot
  {\vec z}^{\pm}&=&0
  \end{eqnarray}
where $\vec z^\pm=\vec v\pm\vec b$ are the Els\"asser variables, $\vec
v$ is the fluctuating plasma velocity, $\vec b$ is the fluctuating
magnetic field (in units of the Alfv\'en velocity), $\vec{V_A}={\bf
  B}_0/\sqrt{4\pi\rho_0}$ is the Alfv\'en velocity based upon the
uniform background magnetic field $\vec{B_0}$, $P=(p/\rho_0+b^2/2)$,
$p$ is the plasma pressure, $\rho_0$ is the background plasma density,
$\nu$ is the fluid viscosity (which, for simplicity, we have taken to
be equal to the magnetic diffusivity) and $\vec f^\pm$ represent
forces that drive the turbulence at large scales. It can be shown that
in the limit of small amplitude fluctuations, and in the absence of
forcing and dissipation, the system describes non-interacting linear
Alfv\'en waves with dispersion relation $\omega^\pm(\vec k)=\pm k_\|
V_A$. The incompressibility condition requires that these waves be
transverse.  Typically they are decomposed into shear Alfv\'en waves
(with polarizations perpendicular to both $\vec{B_0}$ and to the
wave-vector $\vec{k}$) and pseudo-Alfv\'en waves (with polarizations
in the plane of $\vec{B_0}$ and $\vec{k}$ and perpendicular to
$\vec{k}$).

Nonlinear interactions (or collisions) between counter-propagating
Alfv\'en wave packets distort the packets,
splitting them into smaller ones until a scale is reached when their
energy is converted into heat by dissipation~\citep{kraichnan_65}. The efficiency of the
nonlinear interaction is controlled by the relative size of the linear
and nonlinear terms in equation (\ref{mhd-elsasser}): The regime in
which the linear terms dominate over the nonlinear terms is known as
{\it weak} MHD turbulence, otherwise the turbulence is called {\it
  strong}. The Fourier energy spectrum of MHD turbulence can be
derived analytically only in the limit of weak turbulence
\citep[e.g.,][]{ng97,newell_nb01,kuznetsov2001,galtier_nnp02,lithwick_g03,galtier_c06,chandran08a,boldyrev_p09}. 
However, it has been demonstrated both analytically and numerically
that the energy cascade occurs predominantly in the plane
perpendicular to the guiding magnetic field
\citep{newell_nb01,galtier_nnp02,shebalin83,maron_g01}, which ensures
that even if the turbulence is weak at large scales it encounters the
strong regime as the cascade proceeds to smaller scales. Although weak
turbulence may exist in some astrophysical systems
\citep[e.g.,][]{bhattacharjee_n01,saur_etal02,melrose06,chandran08a},
magnetic turbulence in nature is typically strong, for which an exact
analytic treatment is not available. In this case, high-resolution,
well-optimized numerical simulations play a significant role in
guiding our understanding of the turbulent dynamics. This provides the motivation 
for the present work.

The ideal MHD system conserves the Els\"asser energies
$E^+=\frac{1}{4}\int (z^+)^2 d^3 x$ and $E^-=\frac{1}{4}\int (z^-)^2
d^3 x$ (equivalently, the total energy $E=E^v+E^b=\frac{1}{2} \int
(v^2 + b^2)d^3 x=E^+ + E^-$ and the cross-helicity $H^C=\int ({\bf
  v}\cdot {\bf b}) d^3 x= E^+ - E-$ are conserved). The energies $E^+$
and $E^-$ cascade in a turbulent state toward small scales due to the
nonlinear interactions of oppositely moving $\vec{z^+}$ and
$\vec{z^-}$ Alfv\'en packets. MHD turbulence is called {\em balanced}
when the energies carried by oppositely moving fluctuations $E^\pm$
are equal, and it is called {\em imbalanced} when they are not the
same. MHD turbulence in nature and in the laboratory is typically imbalanced. 
For instance, this is the case when the turbulence is generated by
spatially localized sources, as is the case in the solar wind where
more Alfv\'en waves propagate away from the Sun than towards it. 
The independent conservation of the two Els\"asser energies (compared
to only one conserved energy in hydrodynamics) has a profound
consequence for the MHD
dynamics~\citep[e.g.,][]{dobrowolny80,grappin_pl83,meneguzzi96,boldyrev_05,boldyrev_06,mason_etal06,zhou2004,matthaeus_etal08,perez_b09,boldyrev_etal09}.  

In this work we present the results of a series of direct numerical
simulations of MHD and Reduced MHD for balanced and moderately
imbalanced turbulence and investigate how the scalings of the
Els\"asser spectra behave as the Reynolds number is 
increased. We also present the first high-resolution direct comparison
of simulations of MHD vs RMHD turbulence, demonstrating that the
latter model completely captures the turbulence dynamics of strong MHD
turbulence at roughly half the computational cost of a full MHD
simulation.

This paper is organized as follows. In section ~\ref{sec:pheno} we
briefly describe the most recent phenomenological efforts to
understand scaling laws in MHD turbulence, particularly in the
imbalanced case. In section ~\ref{sec:numerics} we describe the
numerical set up and the parameter regime for our simulations. In section
~\ref{sec:spectrum} we show measurements of the energy spectrum from a
series of numerical simulations with varying Reynolds numbers. In
section~\ref{sec:alignment} we show measurements of scale-dependent
dynamic alignment and establish its relation with the $-3/2$ scaling
of the energy spectrum. In section ~\ref{sec:kolmogorov} we discuss
the approach to the universal regime and the universality of
Kolmogorov's constant in MHD. We show that dynamic alignment
introduces a new robust scale-dependent quantity that enters the
definition of the energy spectrum and uniquely sets the Kolmogorov
constant. We propose that this new quantity is a consequence of
cross-helicity conservation. Finally, in section ~\ref{sec:discussion} 
we discuss our results.

\section{MHD turbulence phenomenology}\label{sec:pheno}
For strong MHD turbulence, Goldreich \& Sridhar \cite{goldreich_s95}
argued that the pseudo-Alfv\'en modes are dynamically irrelevant for
the turbulent cascade (since strong MHD turbulence is dominated by
fluctuations with $k_\perp \gg k_\|$, the polarization of the
pseudo-Alfv\'en fluctuations is almost parallel to the guide field and
they are therefore coupled only to field-parallel gradients, which are
small since $k_\| \ll k_{\perp}$). If one filters out the
pseudo-Alfv\'en modes by setting $\vec z^\pm_\|=0$, it can be shown
that the resulting system is equivalent to the Reduced MHD model:
\begin{eqnarray}
  \(\frac{\partial}{\partial t}\mp\vec V_A\cdot\nabla_\|\)\vec
  z^\pm+\left(\vec z^\mp\cdot\nabla_\perp\right)\vec z^\pm =
  -\nabla_\perp P\nonumber\\ +\nu\nabla^2\vec z^\pm +\vec f_\perp^\pm,
  \label{rmhd-elsasser}
\end{eqnarray}  
We note that in RMHD the fluctuating fields have only two vector
components, but that each depends on all three spatial
coordinates. Moreover, because the $\vec z^\pm$ are assumed incompressible
($\nabla\cdot\vec z^\pm=0$), each field has only one degree of freedom
which is more commonly expressed in terms of stream functions in the
more standard form of the RMHD equations~\cite{kadomtsev_p74,strauss_76}.

Conservation of both the Els\"asser energies means that once an
imbalance has been created it cannot be destroyed by the MHD
dynamics. It is also well known that decaying MHD turbulence, affected
only by the dissipation, becomes increasingly more imbalanced with
time \citep[e.g.,][]{dobrowolny80,zhou2004,matthaeus_etal08}. 
Several analytic and numerical studies have shown that imbalance is
also an inherent property of \emph{driven} MHD turbulence even if the
turbulence is forced without introducing a net imbalance at the
largest scales -- the turbulent domain spontaneously fragments into
local imbalanced domains where the cross helicity is either positive
or negative
\citep{grappin_pl83,meneguzzi96,zhou2004,boldyrev_05,boldyrev_06,
mason_etal06,perez_b09,boldyrev_etal09}. 

In imbalanced domains, the directions of 
the magnetic and velocity fluctuations are not independent, rather, they are 
either aligned or counter-aligned to a certain degree~\footnote{A simple geometrical 
consideration shows that a given degree of imbalance sets an upper boundary on 
the allowed alignment angle between the magnetic and velocity vectors. The 
larger the imbalance $z^+/z^-$, the smaller the allowed angle between magnetic 
and velocity fluctuations. The opposite is, however, not necessarily true.}. The organization of such a domain is the following: 
the {\em directions} of both the magnetic and velocity fluctuations vary within a small angle (comparable to the alignment angle) 
throughout the domain, while their amplitudes change predominantly in the direction normal to their polarizations. Such
positively and negatively aligned domains appear to be the building 
blocks of MHD turbulence, whether it is balanced overall or not. 

The origin of such domains can be qualitatively understood 
from the conservation of energy and cross-helicity in an ideal MHD system. When small dissipation is present and the system is unforced, it can be argued that energy decays faster than cross-helicity. This selective decay would eventually lead to Alfv\'enization of the flow, that is, to progressively stronger alignment (or counter-alignment, depending on the initial state) between the directions of the magnetic and velocity fluctuations, e.g., \cite{dobrowolny80,matthaeus80,grappin_pl83,biskamp_03}. In a perfectly aligned (counter-aligned) state either ${\bf z}^+$ or ${\bf z}^-$ is identically zero, and the  nonlinear interaction vanishes. In a driven state, characterized by strong nonlinear interaction and a constant energy flux over scales, the alignment cannot be perfect. Rather, it turns out that alignment depends on the scale, the smaller the scales the better the alignment. Below we will demonstrate this phenomenon in numerical simulations. From a more qualitative point of view, one can argue that whenever a partly aligned domain appears, nonlinear interaction inside such a domain gets reduced, and its evolution time increases compared to non-aligned domains. Therefore aligned domains persist longer, which explains the tendency of a turbulent flow to exhibit such self-organization. These aligned domains 
are the domains where the essential energy of the turbulence is contained, and
they are typically well seen in numerical simulations. Solar wind observations also show that globally balanced turbulence is made up of locally imbalanced patches at all scales~\cite{podesta09,podesta_b10a,chen12}.

In the aligned or imbalanced domains, the Els\"asser
energies are unequal, and one can ask whether their spectra have to be the
same. This raises questions of whether MHD turbulence is universal and
scale-invariant. Indeed, if imbalanced domains have different spectra
that depend on the degree of imbalance, their superposition may not
have a universal scaling.

Phenomenological treatment of strong imbalanced MHD turbulence is
complicated by the fact that one can formally construct two time
scales for the nonlinear energy transfer: The times of nonlinear
deformation of the $z^\pm$ packets at some spatial scale $\lambda$ are
$\tau^\pm\sim \lambda/z_\lambda^\mp$, which can be significantly
different in the case of strong imbalance,
\citep[e.g.,][]{dobrowolny80, matthaeus_etal08}. In recent years,
several phenomenological models attempting to accommodate this
difference have been proposed. However, the theories have generated
conflicting predictions because they use different assumptions
regarding the physics of the nonlinear energy cascade. For example,
the theory by Lithwick {\it et al.}  \cite{lithwick_gs07} concludes
that in the imbalanced regions the Els\"asser spectra have the
scalings $E^+(k_{\perp})\propto E^-(k_{\perp})\propto
k_{\perp}^{-5/3}$; the same spectra were also suggested by Beresnyak
and Lazarian \citep{beresnyak_l08}. The theory by Chandran
\cite{chandran08} proposes that the spectra of $E^+(k_{\perp})$ and
$E^-(k_{\perp})$ are different depending of the degree of imbalance,
while the theories by Perez and Boldyrev \citep{perez_b09} and Podesta
and Bhattacharjee \citep{podesta_b10b} find that the spectra of
$E^+(k_{\perp})$ and $E^-(k_{\perp})$ have different amplitudes but
the same scalings $E^+(k_{\perp})\propto E^-(k_{\perp})\propto
k_{\perp}^{-3/2}$.

One would expect that numerical simulations could clarify the
picture. However, the first numerical simulations of strongly
imbalanced MHD turbulence
\citep[e.g.,][]{rappazzo_etal07,beresnyak_l08,perez_b09} also produced
conflicting results regarding which power law $E^\pm$ should
follow. The conflicting numerical findings apparently reflect the fact
that imbalanced MHD simulations require significantly more
computational effort compared to the balanced cases
\cite{perez_b10_2}. This happens since in the imbalanced domains the
nonlinear interaction is depleted and the Reynolds and magnetic
Reynolds numbers are reduced. This can be formally seen from the fact
that, in a strongly imbalanced domain with $z^+ \gg z^-$ , the $z^+$
field is advected by a low-amplitude $z^-$ field, and therefore $z^+$
becomes directly affected by the dissipation at smaller wave-vectors
(compared with the balanced case), which reduces its inertial
interval. Now, $z^-$ is advected by a strong $z^+$, but $z^+$ is
significantly affected by the dissipation, so the inertial interval of
$z^-$ becomes spoiled as well.

In order to produce large inertial intervals simultaneously for both
Els\"asser fields when strongly imbalanced domains are present in the
flow, one therefore needs to have a significantly higher Reynolds number
as compared to the balanced case. However, as one increases the
Reynolds number, one needs to increase the numerical resolution in
order to appropriately resolve the small scales and to make sure the
numerical run is stable. Therefore, the larger the imbalance, the
larger the numerical resolution required to describe correctly the
Els\"asser spectra.  Fortunately, it has been argued that Reduced MHD
can be used to investigate the universal properties of MHD turbulence,
which offers the advantage that an RMHD simulation can be achieved at
half the cost of an MHD simulation.

\section{Numerical Setup}\label{sec:numerics}

We solve the MHD equations (\ref{mhd-elsasser}) and their RMHD
counterpart (\ref{rmhd-elsasser}) in a periodic, rectangular domain
with aspect ratio $L_{\perp}^2 \times L_\|$, where the subscripts
denote the directions perpendicular and parallel to $\vec{B_0}$,
respectively. We set $L_{\perp}=2\pi$, $L_\|/L_\perp=6$ or $10$ and
$\vec{B_0}=5\vec{e_z}$.  A fully dealiased 3D pseudo-spectral
algorithm is used to perform the spatial discretization on a grid with
a resolution of $N_{\perp}^2\times N_\|$ mesh points. We note that the
domain is elongated in the direction of the guide field in order to
accommodate the elongated wave-packets and to enable us to drive the
turbulence in the strong regime while maintaining an inertial range
that is as extended as possible (see \cite{perez_b10}). This is a physical requirement that should be satisfied no matter what model system, full MHD or reduced MHD, is used for simulations. 

In the case of reduced MHD though, when the $z^{\pm}_\|$ components are explicitly removed, the resulting system (\ref{rmhd-elsasser}) is invariant with respect to  simultaneous rescaling of the background field $B_0$ and the field-parallel spatial dimension of the system, if one neglects the dissipation terms. Therefore, for any strength of the background field $B_0\gg 1$, one can rescale the field to $B_0=1$ and the field-parallel box size to $L_\|=L_\perp$, that is, conduct the simulations in a cubic box. We should note however that the dissipation terms in (\ref{rmhd-elsasser}) are not invariant and they should be changed accordingly under such rescaling.

To save on computational cost we have reduced the field-parallel numerical
resolution for some simulations, i.e., the numerical grid is
anisotropic with $L_{\|}/N_{\|}> L_{\perp}/N_{\perp}$. This is
appropriate since  the energy cascade proceeds
much faster in the field-perpendicular direction, and the energy spectra decline relatively slowly in the field-perpendicular direction and relatively fast in the field-parallel direction. Energies at large
$k_\|$ are therefore reduced and a lower field-parallel resolution is not
expected to alter the behavior of the spectra in the inertial
interval. An isotropic resolution with the value imposed by the field-perpendicular dynamics would therefore be wasteful.

We should however caution that a reduced resolution (or, equivalently,
unreasonably high Reynolds number for a given resolution) may
contaminate the {\em dissipative} physics, even if the inertial
interval is unaffected. For example, if the precise scaling behavior
in the dissipation interval is of interest, as is the case for
extended scaling laws such as the dynamic alignment angle, somewhat
smaller Reynolds numbers may need to be chosen. As a general rule,
whether the numerical simulations are conducted to investigate the
inertial or the dissipation interval, a resolution study must be
performed in order to establish the optimal Reynolds number for a
given task. In particular, it has to be verified that increasing the
numerical resolution while keeping the physical parameters such as
Reynolds number, forcing mechanism, etc. unchanged does not affect the
studied spectra, e.g.,~\citep{mason_etal11}. This point will be
illustrated below in the balanced case.

The turbulence is driven at the largest scales by colliding Alfv\'en
modes~\footnote{Turbulence may also be driven by driving ${\bf v}$ or
  ${\bf b}$ fluctuations at large scales; this does not affect the
  inertial interval, see~\cite{mason_cb08}.}. We drive both Els\"asser
populations by applying statistically independent random forces
$\vec{f^+}$ and $\vec{f^-}$ in Fourier space at wave-numbers
$2\pi/L_{\perp} \leq k_{\perp} \leq 2 (2\pi/L_{\perp})$, $k_\| =
2\pi/L_\|$. The forces have no component along $z$ and are solenoidal
in the $xy$-plane.  All of the Fourier coefficients outside the above
range of wave-numbers are zero and inside that range are Gaussian
random numbers with amplitudes chosen so that $v_{rms}\sim 1$. The
individual random values are refreshed independently on average
approximately $10$ times per turnover of the large-scale eddies. The
variances $\sigma_{\pm}^2=\langle |\vec f^{\pm} |^2\rangle$ control
the average rates of energy injection into the $z^+$ and $z^-$
fields. We take $\sigma^+>\sigma^-$ and in the statistically steady
state we measure the degree of imbalance through the parameter
$h=(E^+-E^-)/(E^++E^-)=H^C/E$. Thus $h=0$ corresponds to balanced
turbulence and $h=1$ defines maximally imbalanced turbulence.  Time is
normalized to the large scale eddy turnover time $\tau_0=L_\perp/(2\pi
v_{rms})$. The field-perpendicular Reynolds number is defined as
$Re_{\perp}=v_{rms}(L_\perp/2\pi)/\nu \approx 1/{\nu}$. In order to
accommodate the reduced field-parallel resolution we have also
modified the diffusion operator in equations (\ref{mhd-elsasser}) and
(\ref{rmhd-elsasser}), i.e., we have replaced $\nu \nabla ^2$ with
$\nu (\partial_{xx}+\partial_{yy})+ \nu_\|\partial_{zz}$.

The system is evolved until a stationary state is reached, which is
confirmed by observing the time evolution of the total energy of the
fluctuations.  The data are then sampled in intervals of the order of
the eddy turnover time. All results presented correspond to averages
over 30-150 samples for each run. As shown in Table~\ref{tab:simlist},
we conduct a number of MHD and RMHD simulations in the balanced and
imbalanced regime in order to investigate the scaling of the energy
spectra as the field-perpendicular Reynolds number increases.
\begin{table}
{\centering \caption{Simulation Parameters.}}
\begin{tabular}{ccccccccc}
\hline\hline
Case & Regime & $N_{\perp}$ & $N_{\|}$ & $h$ & $L_{\|}/L_{\perp} $ & $Re_{\perp}$ & $\nu_{\|}$\\
\hline
 RB1a  & RMHD   & 512  & 256  & 0  & 6 & 2400  & $\nu$ \\
 RB1b  & RMHD   & 512  & 512  & 0  & 6 & 2400   & $\nu$ \\ 
 RB1c  & RMHD   & 512  & 512  & 0  & 6 & 1800   & $\nu$ \\ 
 RB2a  & RMHD   & 1024 & 256  & 0  & 6 & 6000  & $2.5\nu$ \\
 RB2b  & RMHD   & 1024 & 1024 & 0  & 6 & 6000  & $\nu$ \\
 RB2c  & RMHD   & 1024 & 1024 & 0  & 6 & 3200  & $\nu$ \\
 RB2d  & RMHD   & 1024 & 1024 & 0  & 6 & 1800  & $\nu$ \\ 
 RB3a  & RMHD   & 2048 & 512  & 0  & 6 & 15000 & $2.5\nu$ \\
 RB3b  & RMHD   & 2048 & 2048 & 0  & 6 & 15000 & $\nu$ \\
 RB3c  & RMHD   & 2048 & 2048 & 0  & 6 & 9000  & $\nu$ \\
 RB3d  & RMHD   & 2048 & 2048 & 0  & 6 & 5700  & $\nu$ \\
 
 RI1  & RMHD   & 512  & 256 & 0.45 & 10 & 2200  & $\nu$ \\ 
 RI2  & RMHD   & 1024 & 256 & 0.5  & 10 & 5600  & $2.5\nu$\\ 
 RI3  & RMHD   & 2048 & 512 & 0.5  & 10 & 14000 & $2.5\nu$ \\ 
 MB1  & MHD    & 512 & 256 & 0 & 10 & 2200 & $\nu$   \\
 MB2  & MHD    & 1024 & 256 & 0 & 10 & 5600 & $2.5\nu$   \\
 MI1  & MHD    & 512 & 256 & 0.5 & 10 & 2200 & $\nu$   \\
 MI2  & MHD    & 1024 & 256 & 0.5 & 10 & 5600 & $2.5\nu$   \\
 MI3  & MHD    & 2048 & 512 & 0.5 & 10  & 14000 & $2.5\nu$  \\ 
\hline
\label{tab:simlist}
\end{tabular}
\caption{Summary of the numerical runs with different numerical resolutions and different Reynolds numbers. There is no particular scheme used to choose the Reynolds number for a given resolution other than to ensure that the studied scaling properties are well demonstrated and the numerical runs are  stable.}
\end{table}

\begin{figure}[!tb]
  \includegraphics[width=\columnwidth]{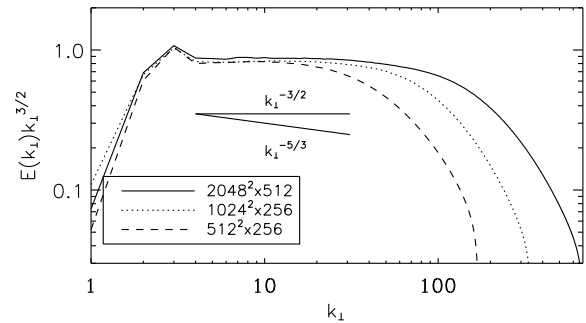}
    \caption{Total field-perpendicular energy spectrum in balanced RMHD as the Reynolds number increases (Cases RB1a, RB2a, RB3a). }\label{fig:balanced_rmhd}
\end{figure}
\begin{figure}[!tb]
\resizebox{\columnwidth}{!}{\includegraphics{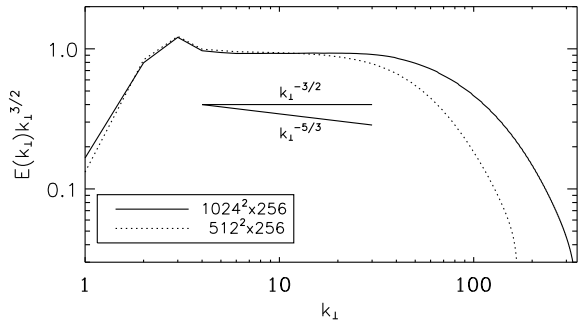}}
    \caption{Total field-perpendicular energy spectrum in balanced MHD as the Reynolds number increases (Cases MB1, MB2).}\label{fig:balanced_mhd}
\end{figure}

\begin{figure}[!tb]
\resizebox{\columnwidth}{!}{\includegraphics{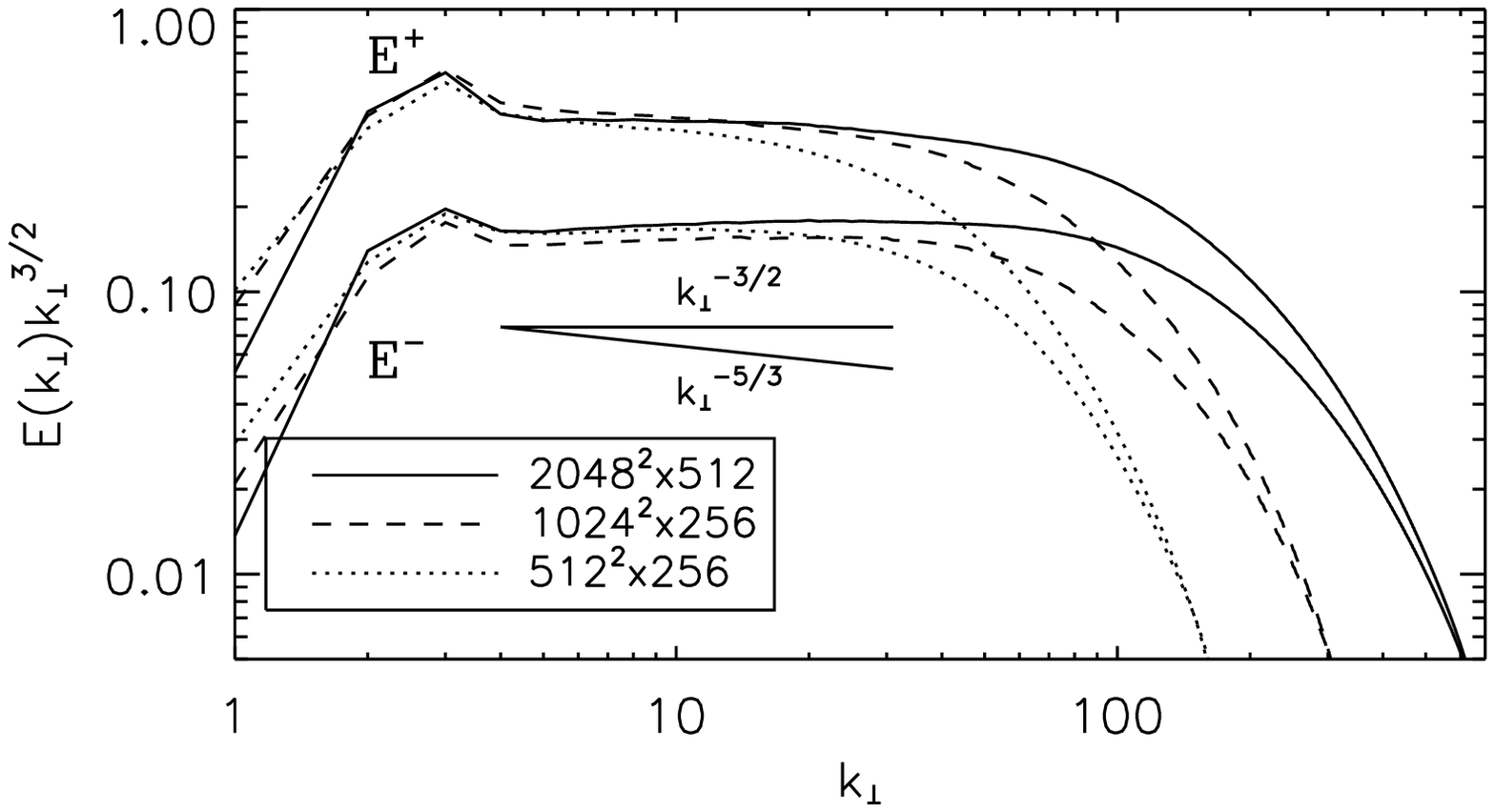}}
    \caption{Energy spectra $E^+(k_{\perp})$ and $E^-(k_{\perp})$  in imbalanced RMHD as the Reynolds number increases (Cases RI1, RI2, RI3). }\label{fig:imbalanced_rmhd}
\end{figure}

\begin{figure}[!th]
\resizebox{\columnwidth}{!}{\includegraphics{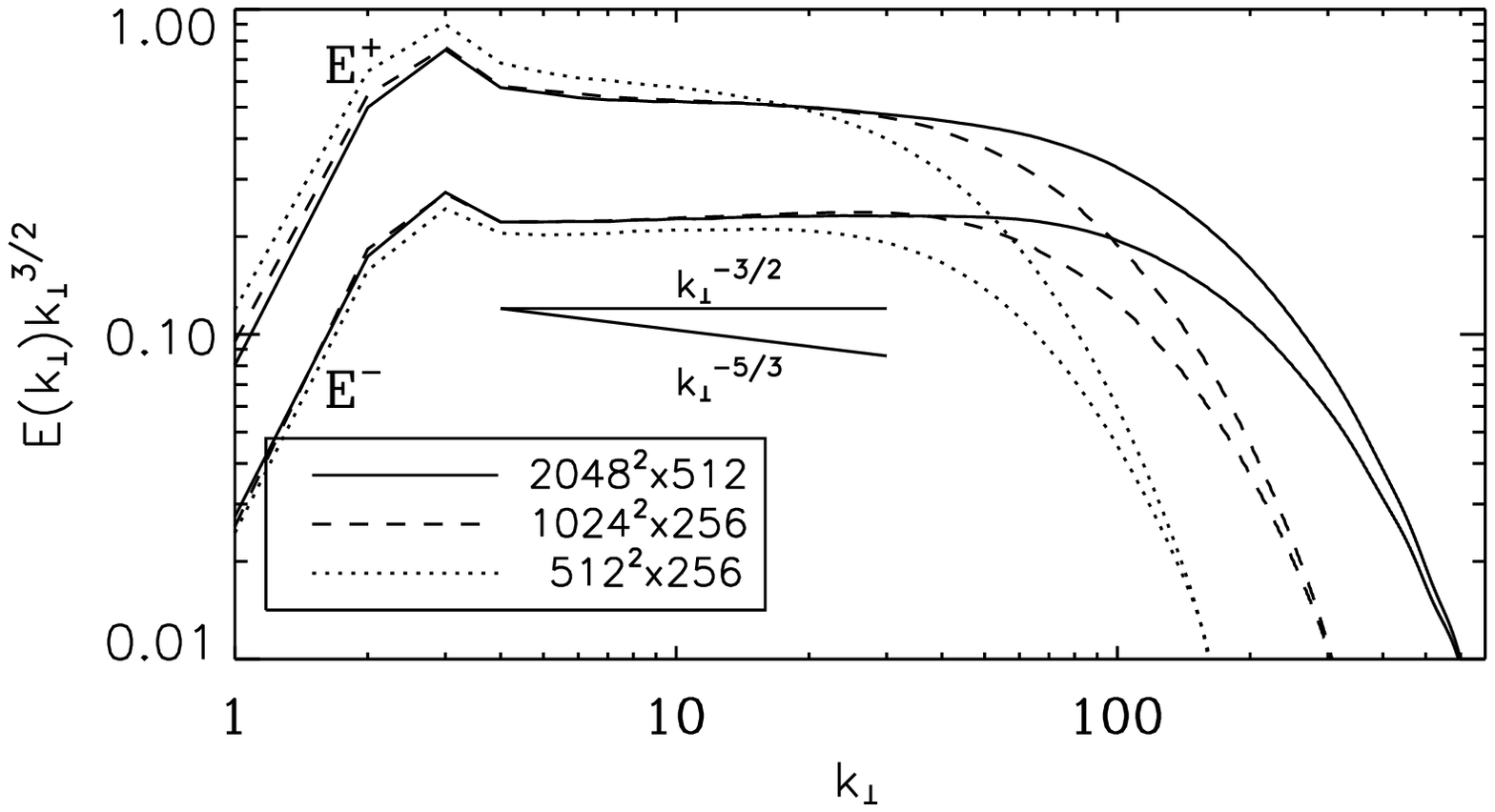}}
    \caption{Energy spectra $E^+(k_{\perp})$ and $E^-(k_{\perp})$  in imbalanced MHD as the Reynolds number increases (Cases MI1, MI2, MI3).}\label{fig:imbalanced_mhd}
\end{figure}

\section{Measurements of the Energy Spectrum}\label{sec:spectrum}

The field-perpendicular energy spectrum is obtained by
  averaging the angle-integrated Fourier spectrum,
  \begin{equation}
    E(k_\perp)=0.5\langle|{\bf v}({\bf k_\perp})|^2\rangle
    k_\perp+0.5\langle|{\bf b}({\bf k_\perp})|^2\rangle k_\perp,
  \end{equation}
  over field-perpendicular planes in all samples.  Identifying the inertial range in numerical simulations with limited
resolution is generally difficult, due to the relatively modest
separation between the forcing and dissipation scales that current
super-computers can afford. For instance, a measurement of the
turbulence spectrum for a single Reynolds number is not enough to
ensure that the simulated turbulence has converged to the asymptotic
universal scaling. Instead, one carries out a set of numerical
simulations with increasing resolution and Reynolds number.  The
spectra are then compensated by the different phenomenological
predictions and the preferred model is distinguished by the best
fit. In Figures \ref{fig:balanced_rmhd} to \ref{fig:imbalanced_mhd}
the inertial range is identified by the flat regions of the spectra
compensated by $k^{3/2}$, which extend further to the right with
increasing Reynolds number (and resolution).

Figures~\ref{fig:balanced_rmhd} and \ref{fig:balanced_mhd} show the
total field-perpendicular energy spectrum $E(k_{\perp})$ in the
balanced regime for the RMHD and MHD cases, respectively.  The RMHD
and MHD spectra are remarkably similar, confirming that the
pseudo-Alfv\'en modes are dynamically insignificant and that the RMHD
approximation is valid.  In both cases the total energy spectrum
remains of the form $E (k_{\perp} )\sim k_{\perp}^{-3/2}$ as the
Reynolds number increases, with the inertial range starting at $k
\approx 4$ and extending up to $k \gtrsim 30$ in the highest Reynolds
number case. In neither RMHD or MHD is there any evidence of a build
up of energy close to the dissipative wave-numbers--often referred to
as a bottleneck effect-- with both spectra falling off smoothly in the
dissipative range.

Figures~\ref{fig:imbalanced_rmhd} and \ref{fig:imbalanced_mhd} show
the field-perpendicular Els\"asser spectra in the imbalanced regime
for the RMHD and MHD cases, respectively. Again the behavior of both
spectra in the RMHD and MHD regimes are very similar. In both cases,
it is seen that while $E^-$ keeps the scaling $E^-(k_{\perp}) \sim
k_{\perp}^{-3/2}$ as the Reynolds number increases, the scaling of
$E^+(k_{\perp})$ is more difficult to pin down. Indeed, both the RMHD
and MHD results for $Re=2200$ yield a steeper spectrum for
$E^+(k_{\perp})$, with an exponent possibly nearer to $-5/3$ than
$-3/2$. However, we believe that there is no real significance to the
value of $-5/3$ here, the exponent is simply steeper than $-3/2$.
Indeed, in both cases, as the Reynolds number is increased
$E^+(k_{\perp})$ appears to flatten, which means that $E^+(k_\perp)$
has not fully established the universal scaling behavior yet. Since
$E^+(k_{\perp})$ and $E^-(k_{\perp})$ are pinned (i.e., converge to
each other) at the dissipation scales and are anchored (i.e., independent of
the Reynolds number) at the driving scales, we postulate that at
sufficiently high $Re$ (where the inertial range is extensive) the
spectra will become parallel in the inertial range and attain the
scaling $E^\pm(k_{\perp}) \sim k_{\perp}^{-3/2}$. Numerical tests of
this prediction must await a significant increase in computational
power.

\section{Measurements of dynamic alignment}\label{sec:alignment}

An important test that can be performed in the presented simulations concerns
the so-called dynamic alignment angle. This angle is {\em
  defined} by the following ratio of the two specially constructed
structure functions \cite{mason_etal06}:
\begin{equation}
  \theta(l) = \frac{\left\langle|\delta \vec v_\perp(\vec
    l)\times\delta \vec b_\perp(\vec
    l)|\right\rangle}{\left\langle|\delta \vec v_\perp(\vec l)||\delta
    \vec b_\perp(\vec l)|\right\rangle},
  \label{daeq}
\end{equation}
where $\delta \vec v_\perp(\vec l)$ and $\delta \vec b_\perp(\vec l)$
are the field-perpendicular velocity and magnetic field increments,
respectively, corresponding to the field-perpendicular scale
separation~${\bf l}$. (We note that in definition (\ref{daeq}) we have
assumed that the angle is small, and hence no distinction between
$\theta(l)$ and $\sin \theta(l)$ is made. Hereafter, by
$\theta(l)$ we will always understand the quantity~(\ref{daeq})).

As proposed in \cite{boldyrev_05,boldyrev_06} the alignment angle $\theta(l)$ has a nontrivial scaling with $l$, which may explain the observed $-3/2$ scaling exponent of the energy spectrum. As
discovered in \citep{mason_etal11}, the scale dependent dynamic  
alignment exists not only in the inertial interval, but it also extends into the dissipation range and is limited only by
the grid size of the numerical scheme.  We will demonstrate that the
alignment angle scaling provides a sensitive test probing the
turbulent cascade deep in the dissipation interval. In particular we
will see that if the simulated dissipation range is under-resolved 
 (e.g., as a result of the use of too large a Reynolds number or strongly
anisotropic resolution), the dynamic alignment can be easily spoiled
at the dissipation scales even if it is present in the inertial
interval.

\begin{figure}[!th]
\resizebox{\columnwidth}{!}{\includegraphics{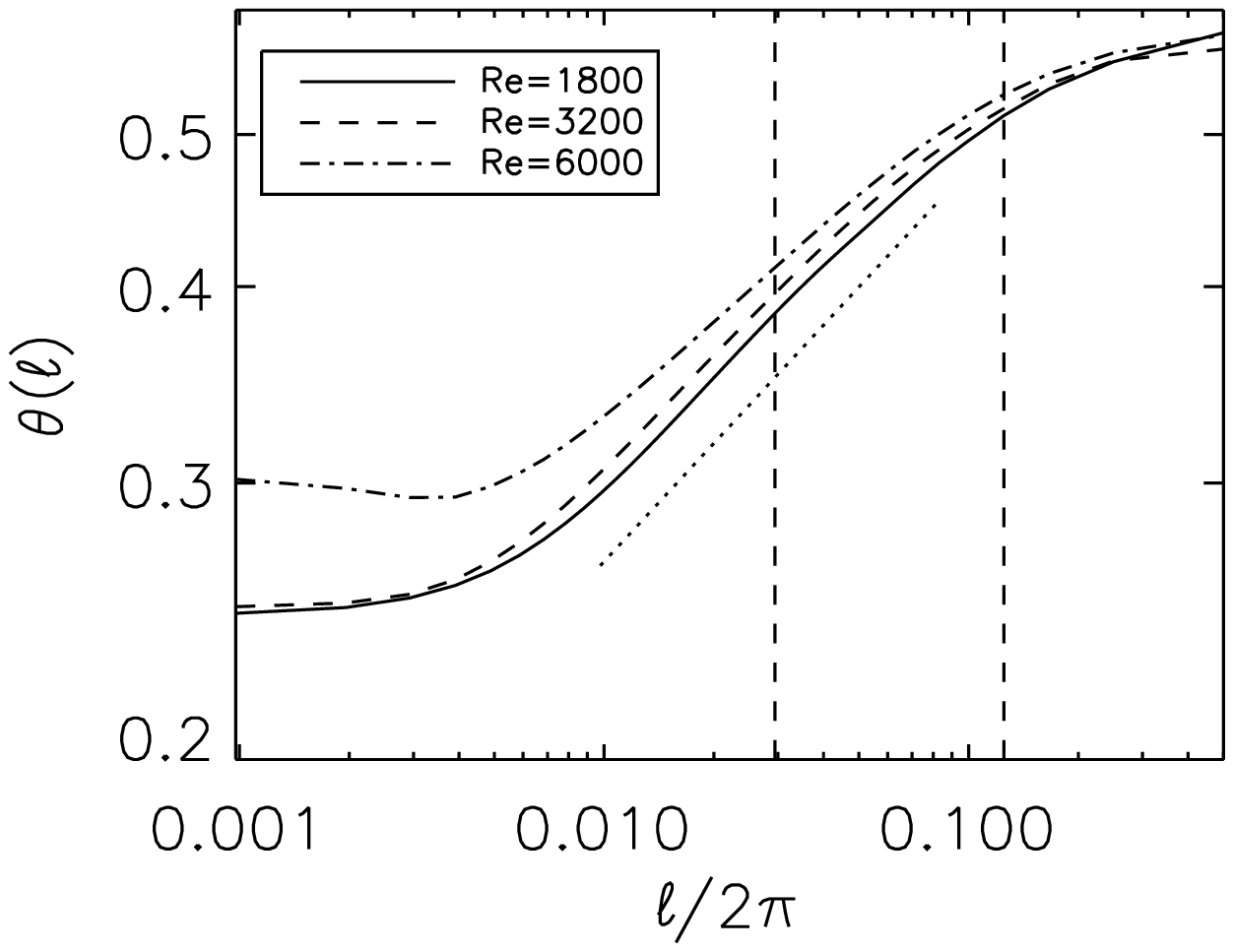}}
\resizebox{\columnwidth}{!}{\includegraphics{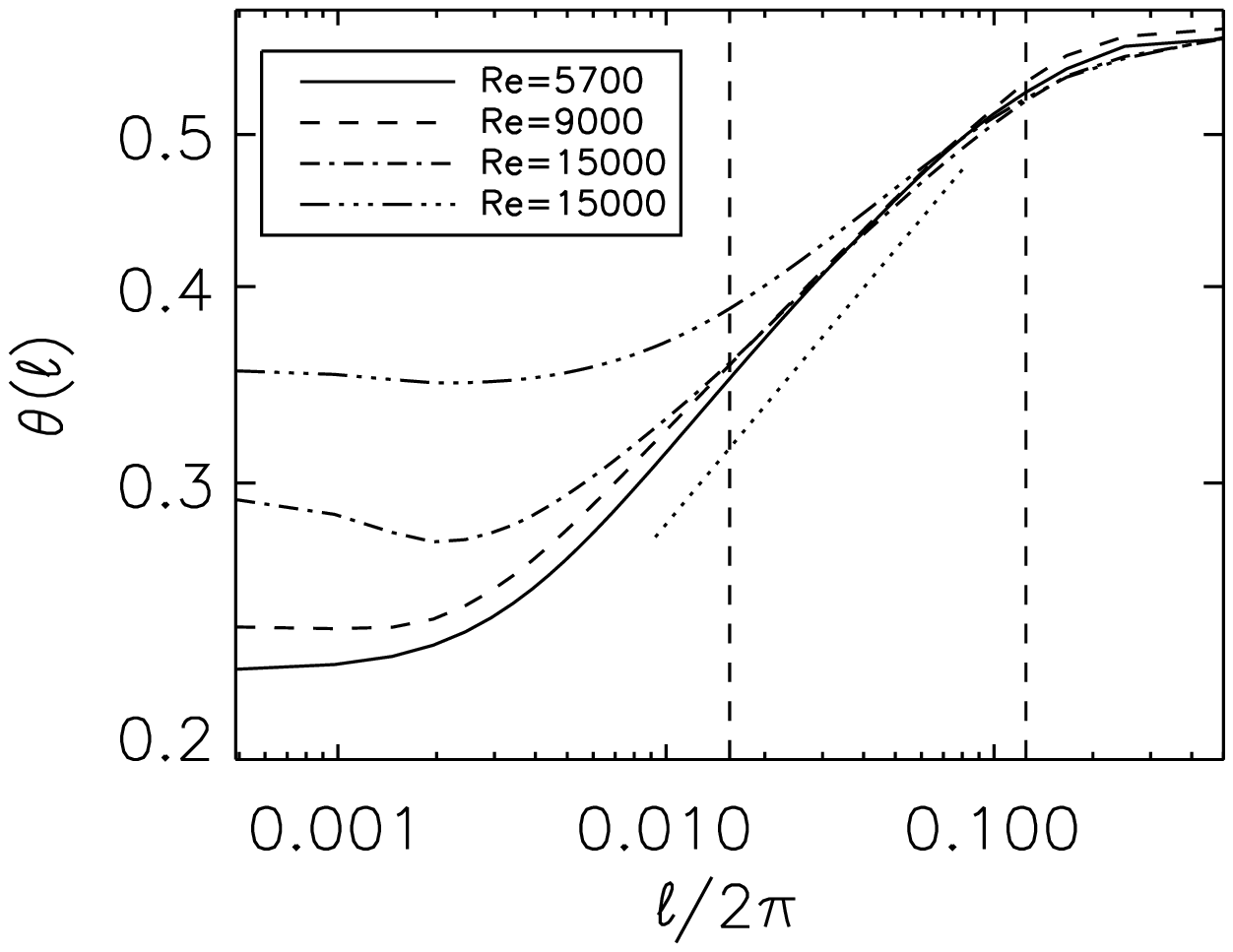}}
    \caption{Measurements of the dynamic alignment angle \eqref{daeq} vs scale $l$
      in balanced RMHD. Upper panel: simulations RB2d (solid), RB2c (dashed), RB2b (dash-dotted) on $1024^3$ mesh points. Lower panel: simulations 
      RB3a (dash-triple-dotted) on $2048^2\times 512$ mesh points, RB3b (dash-dotted), RB3c (dashed), RB3d (solid) on $2048^3$ mesh points at different
      Reynolds numbers. The dynamic alignment scaling extends well
      into the dissipation range, up to scales close to the grid cell
      (roughly $l\sim 3$ grid cells). When the Reynolds number is
      pushed to very high values (so that the dissipation interval
      becomes under-resolved) or the numerical resolution in the
      field-parallel direction is reduced, the alignment-angle scaling
      degrades at small scales. The vertical lines show the
      approximate boundaries of the inertial interval
      (cf. Fig.~(\ref{fig:balanced_rmhd_convergence})). The straight
      dotted line has a slope of $1/4$.}
   \label{fig:alignment_balanced_rmhd}
\end{figure}

The measurements of the alignment angle are presented in
Figure~\ref{fig:alignment_balanced_rmhd}. The first panel shows three
simulations (RB2d,c,b in Table~\ref{tab:simlist}) performed at the
same numerical resolution of $1024^3$ but with different Reynolds
numbers $Re=1800,\, 3200, \, 6000$. Plots for $Re=1800, \, 3200$ show a
remarkable property of the alignment scaling: It extends deep down
into the dissipation region, practically up to the scale of the numerical
discretization, {\em independently} of the Reynolds number (see also
\citep{mason_etal11}). However, this behavior is spoiled if the
Reynolds number is pushed to very high values, at which the
dissipation interval becomes under-resolved. In this case, the scaling
starts to degrade at large wave-numbers, as is seen in the case
$Re=6000$.

\begin{figure}[!th]
\resizebox{\columnwidth}{!}{\includegraphics{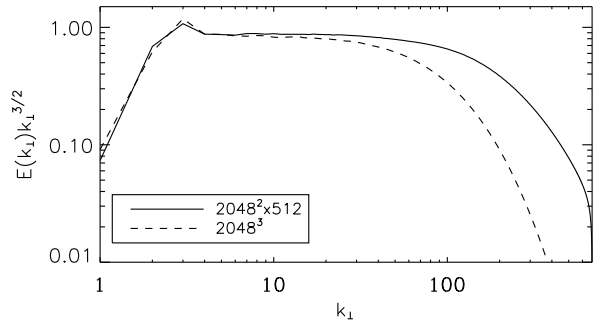}}
    \caption{Energy spectra for runs RB3a (solid) and RB3d (dash). Simulation
    RB3a on $2048^2\times512$ has an unresolved dissipation at the
    expense of longer inertial interval. Simulation RB3d is performed
    at lower $Re$ to capture alignment in the dissipation region, with
    a shorter inertial range.}
    \label{fig:2048_spectra_comparison}
\end{figure}

The alignment scaling is however restored back to its original value
if the numerical resolution is increased to $2048^3$, so that the
dissipation scales become well resolved again. This is seen from
comparison of the plot for RB2b ($1024^3$, $Re=6000$) in the first
panel of Figure~\ref{fig:alignment_balanced_rmhd} with the plot for
RB3d ($2048^3$, $Re=5700$) in the second panel. Further increase of
the Reynolds number in the second panel of this figure demonstrates
that the alignment scaling is stable up to $Re=9000$ (RB3c, $2048^3$),
however, it starts to degrade at large wave-numbers for higher
Reynolds numbers $Re=15000$ (RB3b, $2048^3$), in complete analogy with
the behavior depicted in the upper panel of
Fig.~\ref{fig:alignment_balanced_rmhd} at smaller resolution. The
alignment angle is spoiled even more in the run RB3a ($2048^2\times
512$, $Re=15000$) in the same figure where we simultaneously decrease
the field-parallel numerical resolution, making the dissipation
interval even more under-resolved. Note however, that in both the
first and the second panels of Fig.~\ref{fig:alignment_balanced_rmhd},
the heaviest distortion of the alignment behavior occurs in the
dissipation region, while the inertial interval (approximately
contained between the two vertical lines) is relatively
unaffected. This may explain why an under-resolved dissipation
interval is not manifest in the scaling of energy spectra, as seen in
Figure \ref{fig:2048_spectra_comparison}.

\begin{figure}[!th]
\resizebox{\columnwidth}{!}{\includegraphics{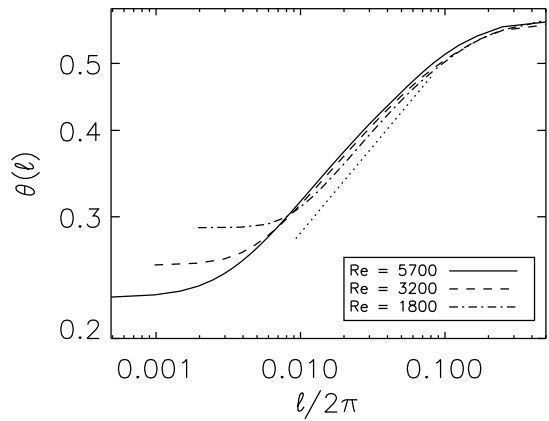}}
\resizebox{\columnwidth}{!}{\includegraphics{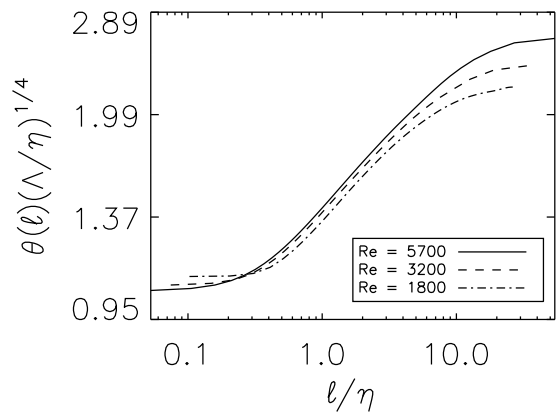}}
    \caption{Measurements of the dynamic alignment angle \eqref{daeq} in balanced RMHD. The frames show numerical simulations with increasing Reynolds number and numerical resolution with properly resolved dissipation ranges (runs RB1c (dash-dotted), RB2c (dashed), and RB3d (solid)). In the lower plot, the scale $l$ is rescaled by the dissipation length (see  section~\ref{energy_scaling} for precise definitions). It can be seen from here that the region of scale dependent dynamic alignment increases as smaller scales are made available by increased numerical resolution. The extent of the alignment region is not limited by the dissipation scale, but rather depends on the grid size of the numerical scheme. The straight dotted line has the slope~$1/4$.}
    \label{fig:alignment_balanced_rmhd_resolved}
\end{figure}

Fig.~\ref{fig:alignment_balanced_rmhd_resolved} shows three well
resolved simulations with numerical resolutions increasing from
$512^3$ to $1024^3$ to $2048^3$. We observe that the scaling interval
of the alignment angle becomes progressively longer and its scaling
index stays close to the predicted value $1/4$ \cite{boldyrev_05} with little or practically no dependence 
on the Reynolds number~\footnote{The actual values of the found numerical slopes are, in fact,  slightly smaller than $1/4$ and, possibly, closer to $0.22$. According to our phenomenological picture this would correspond to the energy spectrum $-1.52$, which is indistinguishable, on a phenomenological level, from the predicted $-1.5$. Possible origins of the discrepancy  may include not large enough Reynolds numbers or/and small intermittency effects.}. This means that we observe a truly universal scaling behavior of the dynamic alignment. The lower panel of Fig.~~\ref{fig:alignment_balanced_rmhd_resolved} shows the same curves where the spatial scale is normalized by the dissipation length. We observe that the flattened parts of the curves at small scales do not overlap under such rescaling, which supports our observation mentioned above that the extent of the scaling interval is not defined solely by the dissipation scale, but rather depends on the numerical discretization step.    

\section{Energy spectrum: Kolmogorov constant and dissipation scale}\label{sec:kolmogorov}
\label{energy_scaling}
For a more complete study of the energy spectrum, one can also
evaluate the amplitude of the spectrum and the dissipation scale for each simulation and verify that they agree with a given phenomenology. Since our spectral
scaling conforms to the phenomenology of Boldyrev
\cite{boldyrev_05,boldyrev_06}, we now study in more detail the
scaling associated with this model.
First, we need to derive the expression for the energy spectrum, which
is done in the following way \cite{boldyrev_05}. The time of
nonlinear interaction at field-perpendicular scale $\lambda$ in this
model is $\tau\sim \lambda/(v_\lambda \theta_\lambda)$, where
$v_\lambda$ denotes the typical (rms) velocity fluctuations,
$\theta_\lambda =\theta_0(\lambda/L_\perp)^{1/4}$ is the
scale-dependent alignment angle between magnetic and velocity
fluctuations, which was studied in the previous section, and $\theta_0$ is the typical alignment angle at the
outer scale (forcing scale) $L_\perp$. The rate of energy cascade is
then evaluated as $\epsilon=v_\lambda^3\theta_\lambda/\lambda$, from
which it follows that
$E(k_\perp)\sim \epsilon^{2/3}(\theta_0/L_\perp^{1/4})^{-2/3}k_\perp^{-3/2}.$
One however notices that the amplitude of the energy spectrum is not
uniquely defined in this equation, since the outer-scale quantities
$\theta_0$ and $L_\perp$ essentially depend on the forcing
routine. This is understood from the following example. Assume that
the large-scale force drives only unidirectional Alfv\'en waves $z^+$,
for which ${\bf v}$ is perfectly aligned with ${\bf b}$ and
$\theta_0=0$. Then the wave energy will grow without bound, since the
nonlinear interaction leading to the energy cascade and eventual
dissipation at small scales is absent.

Even when a particular forcing routine is specified, the definitions
of the values of $\theta_0$ and $L_\perp$ are still subjective since
they essentially rely on the outer-scale properties of turbulence
rather than on the measurements of the inertial interval. We now
propose that this problem can be remedied in an efficient way. For
that we notice that there exists a well-defined quantity that is
remarkably stable (scale-independent) in the inertial interval:
\begin{eqnarray}
\Lambda^{-1/4}=\theta(l)/l^{1/4},
\label{eq:Lambda}
\end{eqnarray}
where $\theta(l)$ is defined in (\ref{daeq}), see the discussion in
the preceding section. In this definition one can use {\em any} scale
$l$ from the inertial interval or dissipation interval if the
numerical simulations are well resolved. A somewhat simpler rule can
be used in numerical (or observational) studies, where one does not
have to know a priori what scales correspond to the inertial interval
and does not have the luxury of having the plot in Fig.~\ref{fig:alignment_balanced_rmhd_resolved} available. In this case
$l$ in formula (\ref{eq:Lambda}) can be chosen to be the Taylor
micro-scale based on either the magnetic or the velocity fluctuations,
$l=v_{rms}/|\nabla \times {\bf v}|_{rms}$ or $l=b_{rms}/|\nabla \times
{\bf b}|_{rms}$, assuming the magnetic Prandtl number is of order one.
We therefore propose the following normalization of the energy
spectrum:
\begin{eqnarray}
E(k_\perp)=C_k \epsilon^{2/3}\Lambda^{1/6}k_\perp^{-3/2},
\label{eq:en_spectrum}
\end{eqnarray}
where $\Lambda$ is {\em defined} by~(\ref{eq:Lambda}). The scale
$\Lambda$ that is defined solely through the inertial-interval
quantities, incorporates the essential information about the
cross-helical structure of MHD turbulence. It is not uniquely defined by 
the outher scale of the turbulence, rather it also depends on the large-scale driving mechanism.  Therefore, the
inertial-interval energy spectrum is defined by the {\em two}
quantities $\epsilon$ and $\Lambda$, characterizing the energy cascade
rate and the level of cross-helical organization of the flow.  The
presence of the two quantities characterizing the spectrum of MHD
turbulence (as oppose to only one quantity in hydrodynamic
turbulence) is the manifestation of the two conserved quantities
cascading toward small scales in MHD turbulence: energy and
cross-helicity.

We expect that the constant $C_k$ in (\ref{eq:en_spectrum}) may be
``universal," that is, largely independent of the character of the
driving, analogous to the Kolmogorov constant in hydrodynamical
turbulence. This constant can be measured in our simulations in the following way. 
First, we specify $l$ that we use to measure the alignment scale $\Lambda$ in 
(\ref{eq:Lambda}). According to our plots in
Figs. \ref{fig:alignment_balanced_rmhd} and
\ref{fig:alignment_balanced_rmhd_resolved}, we may choose $l=0.07L_\perp$, say,
as a scale belonging to the inertial interval and not yet affected
by the numerical resolution effects. Then, for simulations  RB1a, RB2a, RB3a 
we find $\Lambda=1.34L_\perp,1.41L_\perp,1.48L_\perp$, respectively.

The dissipation rate can be evaluated based on the energy spectrum (\ref{eq:en_spectrum}) as follows: 
\begin{eqnarray}
\epsilon = \int E(k_\|, k_\perp)(\nu k_\perp^2 + \nu_\| k_\|^2)
dk_\|dk_\perp .
\label{epsilon}
\end{eqnarray}
Our numerical results confirm that the integral of $\nu_\| k_\|^2$
leads to a negligible correction to the dissipation rate, and
therefore it can be omitted, and we can use the field-perpendicular
spectrum $E(k_\perp)=\int E(k_\|, k_\perp) dk_\|$. Then, for simulations 
RB1a, RB2a, RB3a we find: $\epsilon=0.15,0.15,0.16$.

\begin{figure}[!tb]
\resizebox{\columnwidth}{!}{\includegraphics{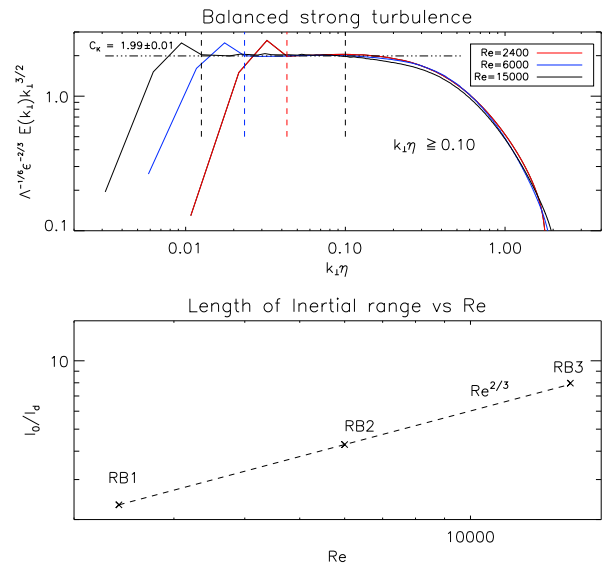}}
    \caption{Upper panel: Total field-perpendicular energy spectrum
      $E(k_{\perp})$ in balanced RMHD turbulence for different
      Reynolds numbers (Cases RB1a, RB2a, RB3a). The wave-number is
      normalized by the dissipation scale (\ref{eta}) and the energy
      is compensated by
      $\Lambda^{-1/6}\epsilon^{-2/3}k_\perp^{3/2}$. The rescaled
      curves collapse onto each other (up to the forcing scale)
      revealing the universal functional form of the energy
      spectrum. Lower panel: The scaling of the
      length of the inertial interval with the Reynolds number. Good
      agreement with the phenomenological model (\ref{eq:en_spectrum}, \ref{eta}) is
      observed.}\label{fig:balanced_rmhd_convergence}
\end{figure}

The dissipation scale can be found (or defined) based on the energy spectrum. 
Omitting the dimensionless constants, we then accept, by definition,
\begin{eqnarray}
\eta = \epsilon^{-2/9} \Lambda^{1/9} \nu^{2/3}.
\label{eta}
\end{eqnarray}
We can demonstrate that our simulations agree with this scaling by
plotting the energy spectra in the balanced case (RB1a,2a,3a) versus
the wave-vector normalized with the dissipation scale~\eqref{eta},
where we measure the dissipation rate directly from the simulations
via (\ref{epsilon}), and the alignment scale from (\ref{eq:Lambda}).   
The top frame in figure~\ref{fig:balanced_rmhd_convergence} shows that
in this case the dissipative region starts around $k\eta\approx
0.1$, independent of the Reynolds number. The extent of the inertial range, defined as the ratio between
the scale $l_0$ at the beginning of the inertial range (from
figure~\ref{fig:balanced_rmhd}, $k = 4$ and hence $l_0 \approx
L_\perp/8$) and the dissipation scale $l_d\approx L_\perp/(2k_d)=5\eta L_\perp$, where
$k_d=0.1/\eta$ from figure~\ref{fig:balanced_rmhd_convergence}), 
increases up to one decade in the RB3a case~\footnote{The
  correspondence $kl_k=\pi$ between eddy scale $l_k$ and wave-vector
  $k$ is assumed.}. Note
that with the wave  
vector normalized with the single parameter $\eta$, the {\em whole
  spectra} collapse onto each other, thus providing additional
evidence that the universal functional behavior of the spectrum is
obtained in our simulations.  The lower plot in figure
\ref{fig:balanced_rmhd_convergence} shows that the length of the
inertial range increases as $l_0/l_d\sim Re^{2/3}$, also in good
agreement with the estimate for the dissipation scale~(\ref{eta}).
The ``Kolmogorov constant'' $C_k$ can be evaluated from the upper plot as $C_k\approx 2$.

\section{Discussion}\label{sec:discussion}
We have presented results from state-of-the-art direct numerical
simulations of balanced and imbalanced driven MHD turbulence.  The
simulations are achieved at the extremely large numerical resolution
of $2048^3$ and the longest running time, with many runs spanning
more than a hundred eddy turnover times in the steady state.  The
simulations were performed using two pseudo-spectral codes, one
solving the MHD equations and the other solving the RMHD equations. In
theories and simulations of MHD turbulence, it has long been argued
that RMHD provides a correct and accurate framework for investigating
the universal properties of MHD turbulence both in the weak and strong
turbulence regimes. We have presented a direct comparison of
high-resolution numerical simulations of MHD vs RMHD turbulence using
two independently developed pseudo-spectral codes with identical
parameters. It is shown that in the strong turbulence regime, in both
the balanced and imbalanced state, the energy spectrum of the
Els\"asser variables in MHD and RMHD are in remarkable agreement (for
details of a lower resolution comparison, including the individual velocity and
magnetic spectra and the alignment angle, see
\cite{mason_etal12}). These results are of essential value for MHD
turbulence research, as simulating MHD turbulence can be accomplished
using RMHD codes that generally incur a smaller computational cost.

In the balanced case, the simulated energy spectra of $E^+$ and $E^-$
show a clearly identifiable inertial range, consistent with a slope of
$k_\perp^{-3/2}$ for both $E^+$ and $E^-$. It is observed from Figures
\ref{fig:balanced_rmhd} and \ref{fig:balanced_mhd} that the
compensated energy spectra show a flat region that extends as the
Reynolds number is increased. This is consistent with previous, lower
resolution simulations of strongly magnetized MHD turbulence, e.g.,
\cite{maron_g01,haugen_04,muller_g05,mininni_p07,mason_cb08,chen_11,tobias_etal2011}. In
the imbalanced case, the interpretation of the numerical results is
not as straightforward. Figures \ref{fig:imbalanced_rmhd} and
\ref{fig:imbalanced_mhd} show that the energy spectrum of $E^-$
remains reasonably close to $k_\perp^{-3/2}$, only slightly changing
its overall amplitude for small Reynolds numbers.  As for the $E^+$
spectrum, the compensated spectrum shows a slope slightly steeper than
$-3/2$ which however flattens as the Reynolds number
increases. Another observation from the large Reynolds number
imbalanced numerical simulations is that the spectra of $E^+$ and
$E^-$ are ``anchored'' at large scales and ``pinned'' at the
dissipation scale. From these results we propose that the energy
spectra of $E^+$ becomes asymptotically closer to $k_\perp^{-3/2}$ as
the Reynolds number is increased. Much higher resolutions, exceeding
the capabilities of today's supercomputers, are required to
conclusively demonstrate this conjecture.

Finally, during the refereeing process, our attention was drawn to recent publications by the group of Beresnyak \& Lazarian \cite{beresnyak_l2010,beresnyak_11,beresnyak_12}, in which the authors address issues similar to the ones contained in this paper. Most of the {conclusions} of those papers appear to be at odds with ours (and with similar results or other groups, e.g.,~\cite{maron_g01,haugen_04,muller_g05,mininni_p07,mason_cb08,chen_11}). We however note that the actual {numerical results} presented in \cite{beresnyak_l2010,beresnyak_11,beresnyak_12} agree with ours in the range of scales that we study, while they differ from ours at very large wavenumbers, e.g., $k\gtrsim 50$ in the runs with highest resolution. Beresnyak \& Lazarian suggest that the true inertial interval exists only at these large wavenumbers where they perform their measurements of the scaling relations.  The formal cause of the disagreement of our conclusions with those by Beresnyak \& Lazarian is thus the numerical measurements being performed in essentially different regions of the phase space. 

The question however remains as to what causes the results of the numerical simulations by Beresnyak \& Lazarian to disagree with ours at small, sub-inertial scales. According to our analysis, the answer is the following: the $k$-space intervals on which references  \cite{beresnyak_l2010,beresnyak_11,beresnyak_12} base their conclusions are significantly affected by numerical effects due to the numerical setup they use. It is not appropriate in their simulations to use those intervals for addressing either the inertial or the dissipation regimes. We however note that the dissipation-range dynamics and the behavior of the numerical solution of the MHD equations close to the numerical cutoff is an interesting and not well studied question. It is therefore worth addressing the differences between our simulations and those by \cite{beresnyak_l2010,beresnyak_11,beresnyak_12} in more detail. Since such analysis is not the main objective of the present work, we have presented  the corresponding discussion in the Appendix.

\acknowledgments
This work was supported by the NSF/DoE partnership grant
NSF-ATM-1003451 at the University of New Hampshire, the NSF sponsored
Center for Magnetic Self-Organization in Laboratory and Astrophysical
Plasmas at the University of Chicago and the University of Wisconsin -
Madison, the US DoE awards DE-FG02-07ER54932, DE-SC0003888,
DE-SC0001794, the NSF grants PHY- 0903872 and AGS-1003451, and the DoE
INCITE 2010 Award.  This research used resources of the Argonne
Leadership Computing Facility at Argonne National Laboratory, which is
supported by the Office of Science of the U.S. Department of Energy
under contract DE-AC02-06CH11357. The studies were also supported by
allocations of advanced computing resources provided by the NSF TeraGrid
allocation TG-PHY110016 at the National Institute
for Computational Sciences and the PADS resource (NSF grant OCI-0821678)
at the Computation Institute, a joint institute of Argonne National Laboratory
and the University of Chicago.

\appendix*
\section*{APPENDIX: Numerical study of MHD tubrulence at subrange scales}
\label{sec:appendix}

In this Appendix we comment on the numerical reconstruction of solution of the MHD equations at small scales, that is, scales within the dissipation range and close to the numerical cutoff in $k$-space (the dealiasing cutoff in a pseudo-spectral code). Recent publications by Beresnyak \& Lazarian \cite{beresnyak_l2010,beresnyak_11,beresnyak_12} found that the energy spectrum in this region (roughly corresponding to $k\gtrsim 50$ in their highest-resolution runs) can have a peculiar structure that is inconsistent with the structure and the scaling found in our numerical simulations. Much confusion was created by the suggestion by \cite{beresnyak_11,beresnyak_12} that this high-$k$ region is, in fact, the true inertial interval of MHD turbulence, while the region that is studied in our works (corresponding to $4 \lesssim k \lesssim 30$ in highest-resolution runs) is a ``non-converged" forcing-dominated region.  

It is therefore useful to address the small-scale numerical solution of the MHD equations in more detail and to relate our findings and conclusions to those presented in the recent works by  Beresnyak and Lazarian \cite{beresnyak_l2010,beresnyak_11,beresnyak_12}.  
These references claim that the energy spectral index of MHD turbulence is $-5/3$ and that there is no conclusive evidence for dynamic alignment in the numerical results. 
In discussing what could lead to such (in our opinion, erroneous) conclusions it is useful to distinguish two factors. One is related to differences that arise because the simulations by Beresnyak \& Lazarian that allegedly are identical to ours, in fact are not identical at all because of differences in the details of the numerical setup. The other is related to the methods that are used to analyze the results and, ultimately, support one claim or another. Both play a role in the origin of the disagreement.

First, we concentrate on issues that result from the different setup of the numerical simulations. In our previous publications (e.g.,~\cite{perez_b10_2,mason_etal12}) we have discussed at length those aspects of the simulation design that are essential for accurately capturing the physics of the strong turbulent cascade. It is not necessary to repeat those discussions here, however, it is important to point out that many of the simulations of Beresnyak and Lazarian \cite{beresnyak_l2010,beresnyak_11,beresnyak_12} differ from ours through their choice of numerical hyper-dissipation, significantly smaller viscosities for a given numerical resolution, and a considerably smaller statistical ensemble from which averages are computed.  Each of
these factors is potentially detrimental for the observation of the 
correct scaling behavior.  For example, the measurements of the
alignment angle that are shown in Figure~3 of reference \cite{beresnyak_11} and Figure~2 of reference \cite{beresnyak_12} lead Beresnyak to conclude that dynamic alignment is not present in MHD turbulence as the alignment angle saturates, i.e. flattens as a function of $l$ at small $l$, when the Reynolds number increases. However, those plots exhibit a behavior that is qualitatively similar to that displayed in our Figure~\ref{fig:alignment_balanced_rmhd},  
where insufficient numerical resolution is demonstrated to affect the alignment angle at small scales. It therefore reasonable to conclude that the observed flattening of the alignment angle in the simulations of references \cite{beresnyak_11,beresnyak_12} is an artifact of unresolved dissipation scales and, possibly, part of the inertial-range scales,
rather than a physical effect.

The influence of hyper-dissipation may be similarly assessed
from comparing the energy spectra obtained in our work with the energy spectra 
obtained in, say, reference \cite{beresnyak_11}. Our spectra in Figures ~\ref{fig:balanced_rmhd}, \ref{fig:balanced_mhd}, \ref{fig:2048_spectra_comparison} \& \ref{fig:balanced_rmhd_convergence} exhibit an extended interval with the scaling $k^{-3/2}$, identified as the inertial interval, followed 
by a steep decline, identified as the dissipation range. The spectra in
Figure~2 of reference \cite{beresnyak_11} also show an extended interval with the scaling $k^{-3/2}$ (interpreted in reference \cite{beresnyak_11} as a ``non-converged'' range) 
followed at large wavenumbers by a very short steepening (interpreted in \cite{beresnyak_11} as the ``inertial interval'') and then flattening and ultimate cut-off. In our view, such an interpretation is incorrect; 
the spectral behavior observed in reference \cite{beresnyak_11} close to the dissipation region is not a property of the inertial interval, but rather  
is evidence of the so-called bottleneck effect that is expected when numerical 
hyper-dissipation is present. Indeed, as discussed in references
\cite{frisch_etal2008,cichowlas_2005}, an energy spectrum abruptly
terminated in $k$-space by hyper-dissipation or by other Galerkin-type truncation mechanisms, exhibits an inertial interval followed
by a pseudo-dissipation region (steepening of the
spectrum), then by a partly thermalized region (a rise in the
spectrum), and then by a far dissipation range (ultimate cut-off).
The measurements presented in References \cite{beresnyak_11,beresnyak_12} are consistent with such spectral behavior, which motivates a natural explanation of their results as an inertial interval with the $-3/2$ scaling, modified by a substantial bottleneck effect close to the dissipation scales.  Moreover, a thermalization brought about by sharp termination of the spectrum in the $k$-space tends to decorrelate small-scale fluctuations, which
otherwise would remain strongly aligned throughout the dissipation
interval, cf. our
Figure~\ref{fig:alignment_balanced_rmhd_resolved}. This is also consistent with the significant loss of dynamic alignment at small scales that is observed in references~\cite{beresnyak_11,beresnyak_12}. 

A more detailed comparison of our results can be made with those MHD simulations by Beresnyak~\cite{beresnyak_12} that employ a physical Laplacian dissipation (simulations R8 
\& R9 in \cite{beresnyak_12}). By evaluating the Reynolds numbers for those calculations in the same way that it is done in our work, $Re=v_{rms}L/(2\pi \nu)$ with $v_{rms}\approx 1$, we find that simulation R8, with a resolution $768^3$ mesh points, is performed at the Reynolds number $Re\approx 8000$, while calculation R9 (resolution $1536^3$) is performed at $Re\approx 20000$. According to our results in Figure~\ref{fig:alignment_balanced_rmhd},  in the simulations with a resolution of $1024^3$ mesh points the dissipation interval is under-resolved already at $Re\approx 6000$, while in the $2048^3$ simulations the dissipation interval is under-resolved at $Re\approx 15000$. Thus, the runs R8 \& R9 of reference \cite{beresnyak_12} that are most similar to ours have {\em lower} numerical resolutions while {\em higher} Reynolds numbers. Therefore, they have essentially unresolved dissipation intervals and, possibly, parts of the inertial intervals. 

The lack of resolution at the bottom of the inertial intervals in the simulations R8 \& R9 can also be seen from the alignment-angle curves shown in Figure 2 of \cite{beresnyak_12}. Under the rescaling applied in that figure, the curves should approach each other in the inertial interval, as they do in our Fig.~\ref{fig:alignment_balanced_rmhd_resolved}, lower panel.  In contrast, one can see only a short region of in Figure 2 of \cite{beresnyak_12} (runs R8 \& R9) where the curves approach each other, approximately within the range $20 \lesssim l/\eta \lesssim 40$. Apparently, this is the only piece of the inertial interval that is resolved, and in this interval the scaling exponent of the angle indeed approaches $1/4$, see Figure 3 in reference \cite{beresnyak_12}, as expected according to our results.

We now turn to the second factor that contributes to the differing conclusions drawn by the Beresnyak \& Lazarian group, namely the method of analysis. We recall that the objective is to determine the scaling behaviour within the inertial range. Concerning the energy spectrum, we assess whether the numerical data preferentially supports $E(k_\perp)\propto k_\perp^{-3/2}$ or  $E(k_\perp)\propto k_\perp^{-5/3}$ directly by compensating the numerical data by $k^{3/2}$ and by $k^{5/3}$ in turn. For the correct model, the inertial range then corresponds to the range of scales over which the compensated spectrum is flat. We always find that $k_\perp^{-3/2}$ provides the better fit, with the inertial range starting at $k_\perp\approx 4$ and extending up to $k_\perp\gtrsim 30$ at highest resolution. As the Reynolds number increases, numerical convergence is demonstrated by the fact that this region maintains its amplitude and scaling and increases in extent to larger wavenumbers, see, e.g. our  Figure~\ref{fig:balanced_rmhd}. 

In contrast, Beresnyak~\cite{beresnyak_12} uses an {\em indirect} method to select the preferred spectral exponent. He uses the two phenomenological models that describe the inertial range characteristics to predict the dissipation scales ($\eta$), plots the compensated spectrum as a function of the dimensionless wavenumber $k\eta$, and identifies the preferred model as that which displays the better convergence properties at {\em large} wavenumbers $k\eta$ as the Reynolds number increases.  Figure~1 of Beresnyak~\cite{beresnyak_12} lead him to conclude that it is the $-5/3$ model that displays the better convergence properties at large $k$.  

It can be shown, however, that the convergence at small scales  observed in \cite{beresnyak_11,beresnyak_12} is a simple artifact of the numerical setup adopted in \cite{beresnyak_11,beresnyak_12}, rather than a physical effect. To explain this, we note that 
any discrete numerical scheme solves only the corresponding discrete algebraic equations.  If the numerical setup is done correctly, the numerical solution approximates the physical one independently of the discretization step.  If, however, a special numerical setup is adopted where $\eta$ is rigidly tied to the grid size such that $\eta N$ is kept the same in all runs (as is done in \cite{beresnyak_11,beresnyak_12}), then the numerical solution plotted as a function of $k\eta$ is  always affected by the discretization in the same way, thus consistently reproducing the same small-scale numerical effects that are present in the setup. The convergence at large $k\eta$ is then the convergence among solutions of the given numerical scheme, which should not be confused with the convergence to the physical solution. 
\begin{figure}[!tb]
\vskip3mm
\resizebox{\columnwidth}{!}{\includegraphics{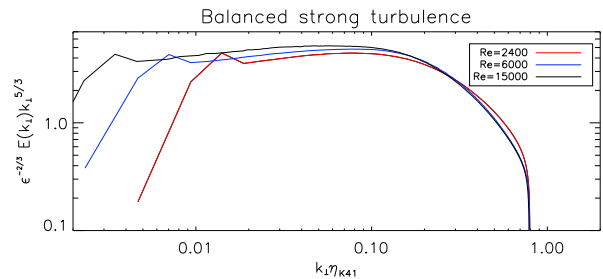}}
    \caption{The spectra presented in Fig.~\ref{fig:balanced_rmhd_convergence} rescaled with a new parameter $\eta_{K41}= \nu^{3/4}\epsilon^{-1/4}$. For this particular choise, $\eta_{K41}$ is proportional to the step of numerical discretization, that is, $N_\perp\eta_{K41} = {\rm const}$. According to our estimate of the onset of the dissipation region in Fig.~\ref{fig:balanced_rmhd_convergence}, the corresponding region in the present plot starts at $k_\perp \eta_{K41}\approx 0.04$, while the numerical dialiasing cutoff that is always imposed at $k_\perp=N_\perp/3$, is seen at $k_\perp\eta_{K41}\approx 0.8$. We observe that the convergence is present in the visinity of the dealiasing cutoff, $k_\perp\eta_{K41} \gtrsim 0.3$, while it is absent in the inertial interval and in most of the dissipation interval.}\label{fig:balanced_rmhd_convergence_k41}
\end{figure} 
\begin{figure}[!tb]
\vskip3mm
\resizebox{\columnwidth}{!}{\includegraphics{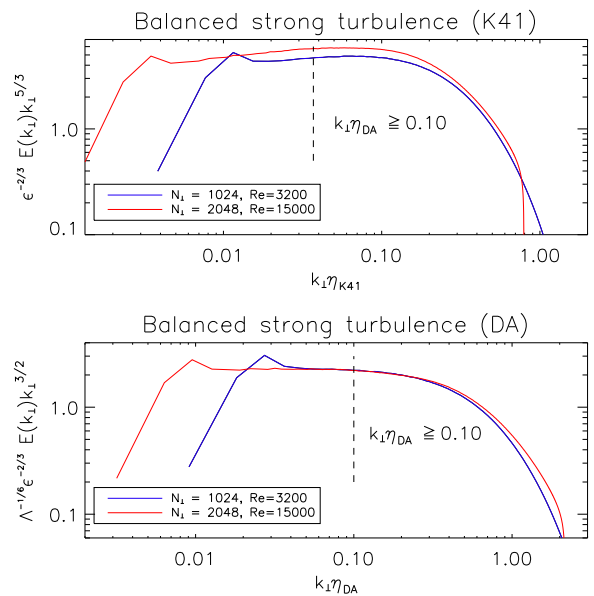}}
    \caption{Comparison of the spectra obtained in numerical runs RB2c and RB3a, rescaled by the Kolmogorov dissipation scale $\eta_{K41} =\nu^{3/4}\epsilon^{-1/4}$ (upper panel) and by the dissipation scale obtained in the dynamic alignment theory $\eta_{DA}=\nu^{2/3}\Lambda^{1/9}\epsilon^{-2/9}$, see~(\ref{eta}) (lower panel). In these runs, the numerical discretization is {\em not} related to the dissipation scale as $N_\perp\eta = \mbox{const}$. With no spurious numerical convergence imposed by the numerical setup, the $-5/3$ model does not fit the data, while the dynamic alignment model shows a good agreement in the inertial interval and in the dissipation range, up to the scales where the numerical effects eventually become significant.}\label{fig:balanced_rmhd_convergence_k41_vs_DA}
\end{figure}

To illustrate this effect in our simulations we replot the spectra presented in  Fig.~\ref{fig:balanced_rmhd_convergence} choosing the Kolmogorov normalization scale $\eta_{K41}= \nu^{3/4}\epsilon^{-1/4}$. Due to a particular choice of viscosities in our runs depicted in Fig.~\ref{fig:balanced_rmhd_convergence}, in this case  $\eta_{K41}$ happens to double every time the resolution decreases by a factor of~$2$, thus ensuring that $\eta_{K41} N_\perp = {\rm const}$, see Fig.~\ref{fig:balanced_rmhd_convergence_k41}.   It is therefore not surprising that  all the curves converge in the vicinity of the numerical dealiasing cutoff corresponding to $k\eta_{K41}\approx 0.8$, while they do not converge in the inertial interval and in the most of the dissipation interval.  A similar, by design,  convergence is present in Fig.~2 of \cite{beresnyak_11} and Fig.~1 of \cite{beresnyak_12}. Such convergence at very small scales is a spurious numerical effect, which does not reflect the convergence of the physical solutions, and which cannot give  preference to any phenomenological model. When the viscosities in different runs do {\em not} conform to the special condition $\eta N_\perp={\rm const}$, the spurious convergence disappears, and the $-5/3$ model does not fit the data, while the $-3/2$ model still provides a good fit in the inertial and dissipation intervals, see Fig.~\ref{fig:balanced_rmhd_convergence_k41_vs_DA}.

We therefore conclude that the numerical simulations by Beresnyak \& Lazarian group    \cite{beresnyak_l2010,beresnyak_11,beresnyak_12} are likely significantly affected by numerical effects at small scales where their measurements are performed. This is notwithstanding the statements made in \cite{beresnyak_l2010,beresnyak_11,beresnyak_12} that the simulations are resolved in those works. These statements, in our opinion, are not supported by the factual numerical data presented in these papers.  Until the effects of hyper-dissipation are better understood and 
numerical convergence is demonstrated in the settings of
\cite{beresnyak_l2010,beresnyak_11,beresnyak_12}, it is hard to assess fully the degree to which the numerical findings of
\cite{beresnyak_l2010,beresnyak_11,beresnyak_12} can be compared with our results, or with similar results of other groups \cite[e.g.,][]{maron_g01,haugen_04,muller_g05,mininni_p07,chen_11}.


\end{document}